\documentclass[11pt,a4paper]{article} 
   
\begin{document}  

\baselineskip 0.55 cm
\begin{flushright}
SISSA 23/01/EP\\
gr-qc/0103073
\end{flushright}

\vskip 1cm
\begin{center}

{\Large{\bf}How strings solve the apparent contradiction between black
holes and quantum coherence \footnote{Talk delivered at the symposium on
``50 years of electroweak physics'' NYU, 27-28 Oct. 2000, in honor of
Alberto Sirlin $70^{th}$ anniversary}}

\vskip 1cm

{\Large Daniele Amati \footnote{e-mail: amati@sissa.it}}

\vskip 1cm

{\it SISSA, 34014 Trieste, Italy}

\end{center}   
    
\vskip 2cm

\begin{abstract}
Computations in the calculable small coupling regime of 
string theories and the general consensus that no new physics has 
to be invoked in continuing to the large coupling (black hole) regime, 
suggest the following picture. Quantum states are not black holes even if energy distributions would suggest so. Black holes appear as macrostates, i.e. with the same procedure that blures the quantum coherence of microstates. It is also discussed how a spacetime description - and thus geometry, causal properties and event horizons - may stem from decoherence in the pregeometric approach represented by string theories. 
\end{abstract}

\newpage

The apparent contradiction between gravity and quantum mechanics, that goes 
under the name of black hole information loss, has to be unambiguously solved
by a consistent quantum theory that contains gravity. Superstring theory is 
thus a serious candidate to clarify the issue, specially since the recognition 
of remarkable coincidences \cite{uno}, suggesting that no new physics has to be invoked in going from a small coupling perturbation regime to a strong coupling black hole one.

By taking seriously this suggestion I will show that a consistent picture emerges \cite{due} in which well defined quantum states - microstates - aren't black holes even if energy densities would classically suggest so. Black holes are generated by the decoherence procedure implied by a mesoscopic (classical) description.

I will first show how this dramatic effect of decoherence is at work in string theories even in the low coupling regime where all properties are perturbatively calculable. Consider a very massive string state, i.e. a high string excitation, in a small coupling regime. This means $(\hbar = c = 1)$

\begin{eqnarray}
  M^2=\frac{n}{\lambda_s^2}~, & ~~n \gg 1~, & 
~~g^2 < \frac{1}{\sqrt{n}} \ll 1~~, \
\end{eqnarray}
 
where $\lambda_s$ is the string length, $g$ the dimensionless string coupling and $n$ the (integer) excitation number. The degeneracy $N_n$ at level $n$ is given by the number of partitions of $n$ in a set of integers $n_i$ such that $n=\sum_{i=1}^{\infty}i n_i$. For large $n$, 
\begin{equation}
N_n\sim \exp [a \sqrt{n}] = \exp[a M \lambda_s]~, \label{N} 
\end{equation}
$a$ being a numerical constant of order one.

The $S$ matrix for whatever initial state may be perturbatively computed and will be unitary. Instead of computing $S$ matrix elements leading to each of the possible final states, well-defined techniques have been developed to compute - again perturbatively - semi-inclusive quantities as spectra of final particles, two particle correlations and so on. This means cross sections for observing one, two or more specific particles with specific momenta and summing over everything else not observed. The total set of semi inclusive quantities stores the whole information on the initial state. 

In ref. \cite{tre} the spectrum in the decay of an arbitrary string state with large mass $M$ was computed at the tree level - weak coupling limit - and shown to give polynomial expressions (in the energy $E$ of a single observed massless particle) that depend, of course, on which is the specific decaying microstate. And it was shown that the spectra, when averaged over the many degenerate states with the same mass $M$, turns out to be the thermal one:
\begin{equation}
F(E) \sim  \frac{e^{-E/T_H}}{1-e^{E/T_H}}~, 
\end{equation}
where $T_H$ is the Hagedorn temperature $T_H = 1/a\lambda_s$, where $a$ is the numerical constant appearing in eq. (\ref{N}). This means that the decoherence of the mesoscopic procedure blures all the information contained in microstate spectra and correlations, leading to the informationless thermal distribution for macrostates. 

It is a well-defined calculation of a consistent quantum theory that shows - baffling some physicists' intuition- how even a specific correlator (i.e. a very limited measure on highly complex final states) may be very different for a microstate or a macrostate. Or, in other words, that the sum due to the non observation of many final state properties, does not imply the decoherence which is  performed by averaging over initial states. This interesting property of string theories is due to their large degeneracy and extreme coherence. Characteristics that are, let me stress, at the very basis of their dual properties and, thus, of their ``raison d'\^{e}tre''. 

The physical picture we discussed was established in the calculable low coupling regime. But if no new physics has to be invoked in the continuation to the strong coupling, where black holes are expected, we recognize how string theory solves the information puzzle. Microstates aren't black holes. Well-defined (and well-prepared) quantum states do not loose coherence and do not jeopardize the quantum character of the whole approach. It is decoherence that throws away information and generates black hole macrostates. The Beckenstein entropy thus measures the degeneracy of microstates of a black hole and does not count black holes. Indeed, the black hole is \emph{the} macrostate and at the macroscopic (incoherent) level it is totally irrelevant how it has been formed. 

The preceding scenario needs however a further understanding of why it is only at the macroscopic level that the geometrical nature of general relativity emerges from string theories. 

String theories contain gravity in the infrared limit: for frequencies much smaller than the string scale, only background fields corresponding to massless states enter in the $\beta = 0$ (no renormalization) condition, thus the metric and its Einstein equation. At the quantum level, however, fluctuations at the string scale will generate all other (massive) background fields that will appear, together with the metric, in a large system of coupled equations. These many non-metric coupled fields, that inhibit a geometrical space time description, are expected to have quickly varying phases, so as to be averaged out in the classical decoherence procedure even for large momenta.
This is not surprising; string theories do not describe the evolution of a string in space time. Space time is not a pre-existing entity described by coordinates $x^{\mu}$; it is a target space, generated by the string, described by the operators $X^{\mu}(\sigma, \tau)$ and it is only at a classical level in which quantum string fluctuations are averaged out, that they emerge as coordinates. 

We thus witness how it is the decoherence procedure that, with the same token, 
 generates the loss of quantum coherence, the emergence of a space time with its geometrical description and thus causal properties, singularities, event horizons and black holes.

I thank Massimo Porrati for interesting discussions during the meeting and for letting me know that ideas similar to mine  have been simultaneously proposed by R. Myers \cite{cinque}.

\end{document}